\documentclass[amsmath,amssymb,aps,twocolumn,prb,longbibliography,floatfix,superscriptaddress]{revtex4-2}
\usepackage{graphicx,bm,times}
\usepackage[utf8]{inputenc}
\usepackage{braket}
\usepackage{amsfonts}
\usepackage{color}

\usepackage[breaklinks]{hyperref}
\hypersetup{
    pdfstartview={FitH},
    colorlinks=true,
    linkcolor=blue,
    citecolor=blue,
    urlcolor=blue
}

\begin{document}

\title{Investigating the magnetism of Ni from a momentum space perspective}

\author{A.~D.~N.~James}
\affiliation{H.~H.~Wills Physics Laboratory,
University of Bristol, Tyndall Avenue, Bristol, BS8 1TL, United Kingdom}

\author{E.~I.~Harris-Lee}
\affiliation{Max Planck Institute of Microstructure Physics, Weinberg 2, D-06120 Halle, Germany}

\author{S.~B.~Dugdale}
\affiliation{H.~H.~Wills Physics Laboratory,
University of Bristol, Tyndall Avenue, Bristol, BS8 1TL, United Kingdom}

\date{\today}

\begin{abstract}
For more than three decades, clear discrepancies have existed between 
spin densities in momentum space revealed by Magnetic Compton scattering experiments and theoretical calculations based on density functional theory (DFT).
Here by making a wide comparison between different theoretical methods, including DFT, DFT combined with dynamical mean field theory, and Hedin's $GW$ approximation, we discover how the magnetic Compton profiles of Ni can be predicted remarkably well. We find that the essential ingredients missing in DFT are (i) local spin fluctuations and (ii) a non-local treatment of electron correlations.
\end{abstract}

\maketitle

\section{Introduction}

The {\it de facto} theoretical technique for predicting the electronic structure of materials is density functional theory (DFT)~\cite{trickey_introduction_1990,ho.ko.64,kohn.99,jo.gu.89,jone.15} which has given crucial insight into the electronic and magnetic properties of an enormous variety of materials.
Yet greater insight can be obtained by going beyond DFT. By applying several different theories, a clearer picture can be formed of the relevant underlying factors which are responsible for the observed phenomena.

Being a rather direct and detailed property of the groundstate many-body wavefunction, the electron momentum density (EMD) serves as a unique and powerful observable  for comparison between theory and experiment.
In this Article we focus on magnetic Compton profiles (MCPs), which are a directly measurable expression of the spin dependence of the EMD \cite{cooper_spin_2000}.
Some of the earliest magnetic Compton profile measurements were performed on the archetypal ferromagnetic metal Ni \cite{sakai_application_1991,timms_spin_1990}, and Ni continues to be the main reference material for calibrating magnetic Compton scattering experiments \cite{billington_2020}.
Ni is not usually considered to be particularly magnetically complex, and DFT predicts the spin moment reasonably accurately. However, DFT calculations do not predict the experimental MCPs of Ni so well, implying that it is fortuitous for DFT to predict such an agreeable spin moment. Despite the attention that has been given to this intriguing problem~\cite{di.du.98,major_refining_2004,ch.be.14a,kubo_electron_2004,james_2021}, no theory has yet been shown to predict the MCPs to within experimental error.

Here, by looking at DFT, DFT with dynamical mean field theory~\cite{held.07,ko.sa.06} (the so called DFT+DMFT method) and Hedin's $GW$ approximation~\cite{hedin_new_1965,hedin_correlation_1999}, we find that a fixed-spin-moment (FSM) `one-shot' $GW$ approximation ($G^0W^0$ \cite{reining_gw_2018}) gives excellent agreement with the experimental MCP data (the prediction is excellent in the sense that it is almost entirely within the standard experimental statistical error at all momenta).
We present evidence that the $G^0W^0$ approximation predicts the shape of the MCPs of Ni so well because it accurately accounts for the beyond-independent-particle electron-electron correlations that lead to occupation distribution smearing and because it improves the description of the delocalized 
magnetization compared to the local spin-density approximation (LSDA) exchange-correlation functional used within DFT. 
The problem is that the $G^0W^0$ calculations worsen experimental agreement compared to LDA by increasing the total spin moment and therefore the total integrated area under the MCPs.
However, we find that this problem can be corrected in a FSM approach with a static external effective magnetic field (or, ostensibly, by combining $GW$ with DMFT) \emph{without} negatively affecting the shape of the MCPs. 
We point out how these results reflect the importance of both the localized and the itinerant magnetism in Ni. We present strong supporting evidence for some of the conclusions made in a recent study of the magnetism of Ni~\cite{sponza_2017}, and we also provide new insight into ways that different (local and non-local) correlations can have an experimentally observable effect.

\section{Background}

\subsection{Magnetic Compton scattering}
The electron momentum density, $\rho(\mathbf{p})$, which is most simply defined as the square modulus of the momentum-space groundstate many-body wavefunction, can be experimentally accessed through x-ray Compton scattering (inelastic scattering of photons by electrons)~\cite{cooper:85,cooper_compton_1997,cooper_x-ray_2004}.
The central principle of this type of experiment is that, within the impulse approximation, the directly measurable double differential inelastic photon-electron scattering cross-section $\sigma_c$ (with respect to measured x-ray energy $\omega$ and solid angle $\Omega$) is directly proportional to a so-called Compton profile, 
\begin{equation}
    \frac{\mathrm{d}^2\sigma_c} {\mathrm{d}\Omega \mathrm{d}\omega} \propto J(p_z),
\end{equation}
where the Compton profile, $J(p_z)$, is the one-dimensional projection of the EMD (i.e. a double integral of two momentum components perpendicular to the scattering vector which is parallel to $p_z$ by convention):
\begin{equation}
J(p_{z})= \iint \left[ \rho_\uparrow(\mathbf{p})+\rho_\downarrow(\mathbf{p}) \right] 
\mathrm{d} p_{x}  \mathrm{d} p_{y}. \label{eq:J}
\end{equation}
In this equation $\uparrow$ and $\downarrow$ denote spin states, $\rho_\uparrow(\mathbf{p})+\rho_\downarrow(\mathbf{p})$ is the total EMD. 
The full three-dimensional EMD can be reconstructed from a relatively low number of Compton profile measurements along various scattering vectors \cite{kontrym-sznajd_fermiology_2009,ketels_momentum_2021}, although detailed information can usually be drawn from the Compton profiles themselves if they are measured for high-symmetry crystallographic directions.
In recent years, Compton scattering has been used to reveal the electronic structure and Fermi surfaces of electronically complex materials such as substitutionally disordered alloys \cite{robarts_2020,billington_2020} and compounds with high vacancy concentrations \cite{ernsting2017vacancies}.
Since photons can only be scattered from the occupied electron momentum states, Compton scattering is sensitive to the Fermi surfaces of metals \cite{dugdale:14}.
Most relevantly, Compton scattering is able to probe the electron correlations within complex materials  \cite{sakurai:95,schulke:01,ruotsalainen2018isotropic,billington2015magnetic,James_2023}. 
Therefore, Compton scattering offers a valuable and complementary perspective on electronic structure and, in particular, a window onto electron correlations in different regimes of composition, temperature and magnetic field from those which other probes can reach. 

If the incident photon beam has a component of circular polarization, the scattering cross-section contains a term which is dependent on the spin of the scatterer. 
This term may be isolated from the charge scattering either by flipping the direction of the sample magnetization or flipping the photon helicity between parallel and antiparallel directions with respect to the scattering vector.
The result is the so-called magnetic Compton profile (MCP)~\cite{sakai_magnetic_1996,cooper_spin_2000}. 
In analogy to the Compton profile, the magnetic Compton profile $J_{\mathrm{mag}}(p_z)$ is the 1D projection of the spin-dependence of the electron momentum density along a chosen scattering vector, given by
\begin{equation}
 \label{eq:J_mag}
  J_{\mathrm{mag}}(p_z)= \iint 
    \left[
      \rho_\uparrow(\mathbf{p})-\rho_\downarrow(\mathbf{p})
    \right]
    \mathrm{d}p_x \mathrm{d}p_y.
\end{equation}

\subsection{Magnetic Compton Profiles of Ni}
The first meaningful measurements of Ni MCPs were made by Timms \textit{et al.} \cite{ti.br.90} and Sakai \textit{et al.} \cite{sakai_application_1991}. 
When convoluted with an appropriate resolution function, the accompanying DFT calculations gave broad agreement with these measurements.
However, the experimental resolution was poor enough to obscure almost totally the fine detail present in the raw (uncovoluted with the experimental resolution) theoretical calculations.
Achieving the highest momentum resolution to date, the experiments of Dixon \textit{et al.}~\cite{di.du.98} revealed the presence of previously obscured finer detail such as periodic (so-called \textit{umklapp}) peaks superimposed on the spectrum. Their results were being independently verified \cite{ka.ku.03} shortly thereafter.
Although the agreement between experiment and theory was generally considered to be ``good'' overall, there were clearly intriguing discrepancies, particularly in the low momentum domain (less than 1.5 a.u.).

The most recent experimental development was the reconstruction of a three dimensional $\rho(\textbf{p})$ from 13 MCPs by Nagao \textit{et al.} \cite{nagao_momentumdensity_2008}.
By making this 3D reconstruction, Nagao \textit{et al.} were able to associate discrepancies between experiment and theory with particular regions of \textbf{\textit{k}}-space, and, beside noting the same disagreement at low momentum, they observed discrepancies at momenta associated with the \textbf{\textit{k}}-space domain at and near to the $X$ point of the Brillouin zone. 
The low momentum region is dominated by the contribution of the most itinerant electrons. 
Following these experiments, there have been multiple attempts to explain the observed discrepancies, which are generally attributed to the inadequate treatment of electron-electron correlations in the widely used DFT exchange-correlation functionals.

%
An alternative to DFT is $GW$, which follows from Hedin's equations \cite{hedin_new_1965,hedin_correlation_1999}.
$GW$ approximations go beyond standard density functional approximations in the sense that the electron-electron correlations are explicitly calculated on a non-local and energy dependent level \cite{reining_gw_2018}.
Kubo was the first to go beyond DFT in calculating the MCPs of Ni, by making a  $G^0W^0$ approximation \cite{kubo_electron_2004}.
Kubo's results showed worse overall agreement with experiment compared to the DFT (LSDA) MCPs, but in several key ways the calculations were limited by the computational resources that were available at the time.
Our new $G^0W^0$ calculations are very different, and this will be revisited in Section~\ref{sec:diagonal_GW}.

Over the last decade, it has been demonstrated that by combining dynamical mean field theory (DMFT)~\cite{me.vo.89,ge.ko.96,ko.vo.04} with DFT (the DFT+DMFT approach \cite{held.07,ko.sa.06}), many of the electronic ground state properties of $d$-block elemental metals, as well as their alloys and compounds, can be better described~\cite{ko.sa.06,held.07,ka.ir.08}. This success comes from capturing all of the local electron-electron interaction vertices within the DMFT framework which enables the prediction of certain phenomena, such as the Mott insulating state~\cite{ge.ko.96}.
DFT+DMFT calculations often have good agreement with measured features from photoemission spectroscopy techniques (PES, and its angle-resolved counterpart, ARPES) which are not described well within DFT.
For Ni there are several of these poorly described features : (i) the dispersionless spectral feature at a binding energy of about 6~eV (known as the ``6~eV satellite''~\cite{gu.ba.77,kakizaki_fluctuating_1994,biermann_first-principles_2003,martensson_investigation_1984,hufner_xps_1972}), (ii) the $3d$-band widths (which were up to $30\%$ narrower than the value obtained from the DFT (LSDA) calculations~\cite{hufner_xps_1972,altmann_enhanced_2000}), (iii) the measured exchange splitting (where both the LSDA and the generalized gradient approximations~\cite{le.ch.91} within DFT overestimate the experimental exchange splitting of the majority and minority states by 0.3~eV, leading to a value that is twice as big as measured)~\cite{ea.hi.78,di.ge.78,hi.kn.79,ea.hi.80}. 
Indeed, DFT+DMFT has now been found to reproduce the exchange splitting and the $6$~eV satellite structure in the valence band~\cite{li.ka.01,br.mi.06,sa.br.12}. 
In addition to this, DFT+DMFT has been used to model the local moment in ambient and Earth-core-like conditions \cite{hausoel2017local} and is also able to predict the temperature dependence of the local moment \cite{li.ka.01,hausoel2017local}, where the ferromagnetic moment is suppressed with increasing temperature (up to the Curie temperature). 

Recently, the directional Compton and magnetic Compton profiles which included the correlation effects from DMFT~\cite{be.mi.12,ch.be.14a} facilitated a discussion of the anisotropy of the electronic correlations of Ni as a function of the on-site Coulomb interaction strength, $U$. 
Those theoretical comparisons with the experimental data led the authors to the conclusion that the theoretical MCPs improved when the local correlations are taken into account. 
However, the DFT+DMFT Ni MCPs calculated by the authors of Ref.~\cite{james2021magnetic} showed that the spin moment depended on $U$ (the spin moment was suppressed with increasing $U$) and that there was no $U$ and $J$ value which gave good agreement between the DFT+DMFT MCPs and the experimental data over all momenta.
The most noteworthy disagreement between the DFT+DMFT MCPs and the experiment was still in the low momentum region.
In particular, DMFT does not strongly alter the shape of the MCPs from the DFT (LSDA) predictions, and therefore full agreement with experiment is not achieved.

\section{Methods}

For these DFT and DFT+DMFT calculations we use the same parameters as are described in Ref.~\cite{james2021magnetic}. 
We perform DFT self-consistent calculations with the {\sc elk} code~\cite{elk} with a full-potential augmented plane-wave plus local orbital (APW+lo) basis~\cite{singh_planewaves_2006}.
For fcc Ni we use a lattice parameter of $a=3.524$~\AA~\cite{WEI20071019},
 the LSDA~\cite{perdew1992} for the exchange-correlation potential and a 20$\times$20$\times$20 Monkhorst-Pack $k$-mesh. 
For comparison with the experimental data, we present FSM DFT MCPs (the spin moment being fixed to the experimental spin moment by a uniform external effective magnetic field). 

For the DFT+DMFT calculations, we construct the Wannier projectors such that the Ni-$d$ states, which are completely within the chosen correlated energy window of ${[-10, 3]}$~eV, are captured. 
We calculate the MCPs for the DFT+DMFT method using the full rotationally invariant form of the interacting Hamiltonian parameterized by $U=2.0$~eV and Hund's coupling ${J=0.9}$~eV, and the spin-polarized around-mean-field (AMF) double-counting term~\cite{pe.ma.03,cz.sa.94}. We use the {\sc TRIQS} library~\cite{triqs} to implement the DMFT cycle along with the {\sc TRIQS}/DFTTools~\cite{aichhorn2016} interfacing between the DFT and DMFT, as described in Ref.~\cite{james_2021}.
We obtain the experimental spin moment (0.56$\mu_B$) for these DFT+DMFT parameters. 
The continuous-time quantum Monte Carlo impurity solver is used ({\sc TRIQS}/CTHYB~\cite{seth2016}) with 4.2$\times$10$^{8}$ sweeps and the inverse temperature $\beta$ of $40$~eV$^{-1}$ (290 K). 
We also present the MCPs from a DFT+DMFT calculation where the magnetism was only treated in the DMFT part of the cycle (i.e., non-magnetic DFT is used and the DFT part remains non-magnetic throughout the entire DFT+DMFT cycle). 
We refer to this type of calculation as DFT+DMFT($\sigma$). 
For this DFT+DMFT($\sigma$) calculation we use the same parameters as for the other DFT+DMFT calculation but we calculate the AMF double-counting term from the spin-averaged density matrix. 
We note that all of this means that the spin moment in DFT+DMFT($\sigma$) is only comprised of a local spin moment within the $d$-orbitals treated within the DMFT cycles.
The method we use for obtaining the natural orbitals and occupation distribution for DFT+DMFT is described in Ref.~\cite{james_2021} and Ref.~\cite{james2021magnetic}. 

For the $G^0W^0$ calculations we start from converged DFT LSDA results and use a 10$\times$10$\times$10 $q$-point-mesh (commensurate with the 20$\times$20$\times$20 $k$-point-mesh).
There are various ways to perform $G^0W^0$ calculations, and, additionally, there are different ways to calculate the EMD and Olevano \textit{et al.} have evaluated several methods for obtaining the momentum distribution from the results of $GW$ approximation calculations \cite{olevano_momentum_2012}.
Here we use the $G^0W^0$ implementation that is available as part of the {\sc elk} code~\cite{elk}, which is a Matsubara method.
First we obtain a new Green's function $G$ from the $G^0W^0$ self-energy by solving the Dyson equation.
We then calculate the natural occupation distribution and orbitals \cite{lowdin_quantum_1955} from this Green's function. 
We note that with the Matsubara method implemented in {\sc elk}, negligible numerical error is introduced if the self-energy is numerically evaluated to high enough Matsubara frequency. Here we evaluate it numerically to 2.042 keV. 
We use the same Matsubara temperature as in the DMFT calculation ($40$~eV$^{-1}$). The total spin moment for the Ni $G^0W^0$ calculations is somewhat sensitive to input parameters such as the Matsubara frequency cut-off and temperature, varying between approximately 0.67--0.75~$\mu_B$ (although the shape of the $G^0W^0$ MCP remains relatively robust).
We present results for both the full self-energy matrix ($G^0W^0$) and the diagonal approximation to the self-energy ($G^0W^0$-DIAG) (in which we only keep the diagonal elements of the self-energy in the initial Kohn-Sham basis).
Furthermore, we refer to $G^0W^0$ calculations with an external uniform effective magnetic field (which we choose so that the spin moment is fixed to the experimental value) by the name $G^0W^0$ (FSM).
We note that by doing this we force the moment of the starting point DFT calculation prior to the $G^0W^0$ step to be 0.43 $\mu_{\rm B}$. 

Here we always calculate the spin-dependent EMD from the orbitals and occupation distribution which are obtained at the end of the respective theoretical calculation ($G^0W^0$, DFT+DMFT, or DFT). 
For example, we note that for a converged DFT calculation, the spin-dependent EMD, $\rho_\sigma(\mathbf{p})$, can be defined in terms of the Fourier transformed real-space Kohn-Sham eigenfunctions $\psi^\sigma_{\mathbf{k}, \eta}(\mathbf{r})$: 
\begin{equation}
  \rho_\sigma(\mathbf{p})= \sum_{\mathbf{k}, \eta}n^{\sigma}_{\mathbf{k}, \eta}\left|\int\limits_{V} \exp (-i \mathbf{p} \cdot \mathbf{r}) \psi^\sigma_{\mathbf{k}, \eta}(\mathbf{r}) \mathrm{d} \mathbf{r}\right|^{2}, \label{EMD}
\end{equation} 
where $n^{\sigma}_{\mathbf{k}, \eta}$ are occupation distributions for eigenstate index $\eta$, and spin index $\sigma$. 
Here we use a linear tetrahedron method \cite{ernsting_2014} to evaluate Eqn.~\ref{EMD}.
We use a maximum momentum cut-off of ${16}$~a.u. (because we find the valence electron EMD to be approximately zero beyond this). 

We note that unfortunately a band-by-band resolution of the EMD is not possible in the DFT+DMFT and $G^0W^0$ frameworks due to there no longer being a one-to-one correspondence between the quasiparticle bands and the natural orbital basis used within DFT+DMFT and $G^0W^0$ to calculate the EMD.
However, we can exploit the dual nature of the basis functions (APW+lo~\cite{singh_planewaves_2006}) to provide an approximate guideline for how the localized and delocalized character spin states each affect the total spin momentum density. 
To separate the total EMD into muffin tin and interstitial parts used in the APW+lo framework, we calculate the muffin-tin (spherical harmonic basis function) contribution to the EMD for each atom and then we determine the interstitial contribution from the difference between the total EMD and the sum of all of the atomic muffin-tin parts.

We calculate the MCPs by integrating the spin-dependent EMD along the appropriate scattering vector (Eqn.~\ref{eq:J_mag}).
In explicitly stated cases, we convolute our MCP calculations with a Gaussian function with a full-width-at-half-maximum (FWHM) of $0.43$~a.u. to represent the experimental momentum resolution.
We normalise the total integral of the EMD to be equal to the number of valence electrons. With this normalization, the areas under the MCPs (their total integrals over momentum) are equal to the spin magnetic moment.
This is, in fact, how we calculate the quoted $G^0W^0$ moment.
For our DFT calculations, we find that the spin moment obtained through this (unusual) method is identical to the moment obtained by the (usual) real-space integration method. 

\section{Results}
\label{sec:mcp}

\begin{figure*}[!ht]
  \centerline{\includegraphics[width=0.90\linewidth]{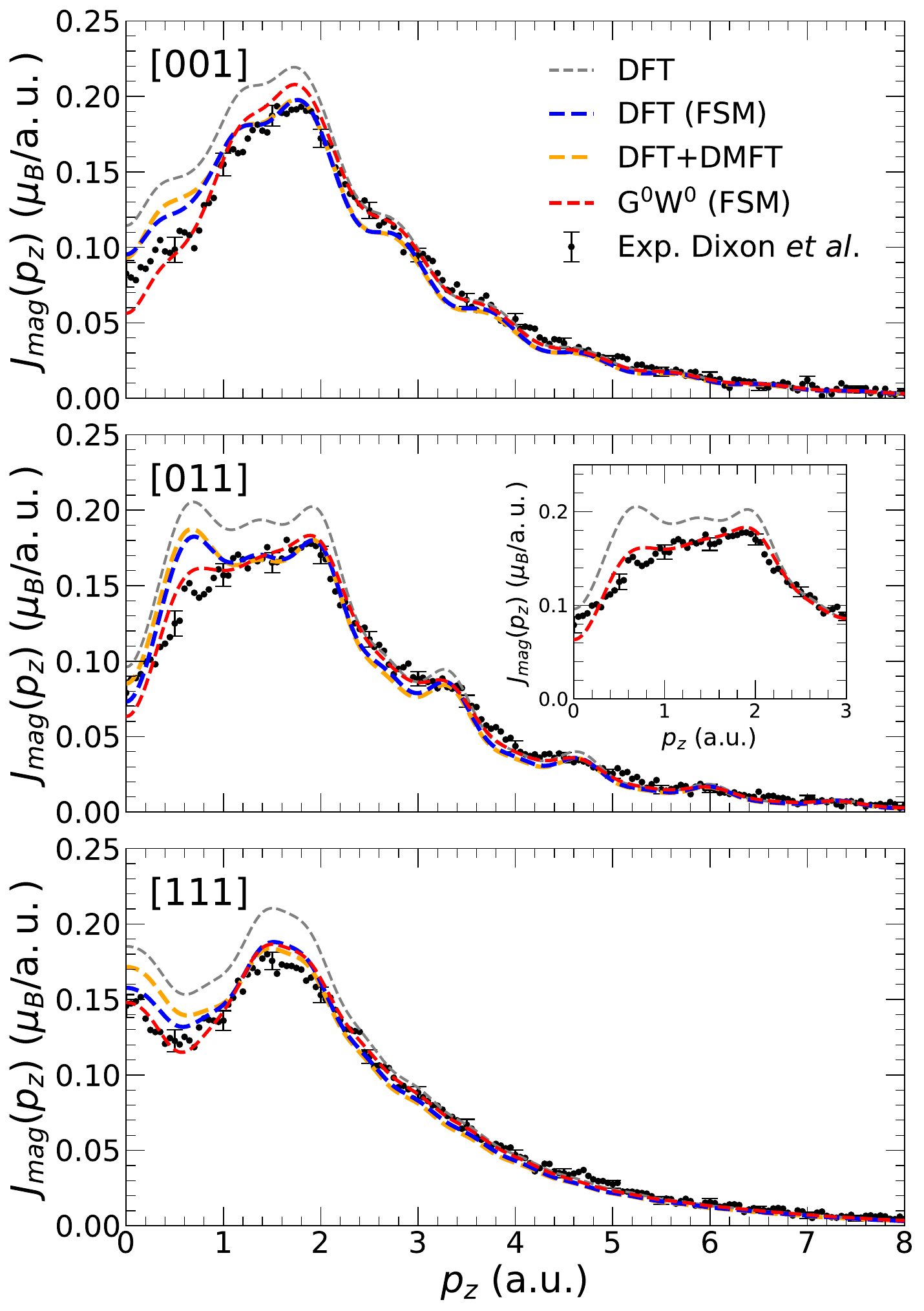}}
  \caption{Comparison of three high-symmetry direction ([001], [011], [111]) experimental Ni magnetic Compton profiles (MCPs)~\cite{di.du.98} (dots with error bars) with our DFT (LSDA), fixed-spin moment DFT (FSM), DFT+DMFT, and $G^0W^0$ (FSM) MCPs. 
  The inset of the [011] MCP puts a spotlight on the comparison between the standard DFT and the $G^0W^0$ (FSM) calculations.
  We show the experimental error bars for every tenth data point. 
  We find that beyond 8~a.u. the experimental and theoretical MCPs are not distinguishable.
  We convolute our calculations with an experimental resolution function.
}
 \label{fig:theory-vs-experiment}
\end{figure*}

We present the theoretical MCPs which have comparable spin moments with the experiment in Fig.~\ref{fig:theory-vs-experiment}.
Immediately, we note how remarkably well our $G^0W^0$ (FSM) calculations compare to the experimental data \cite{di.du.98}, which is vastly superior to the other theoretical MCPs.
In particular, in contrast to the DFT calculation, we find that there is no anomalously large density at $p_z=0.7$~a.u. in the [011] profile, that the repeating (\textit{umklapp}-like) peaks at higher momenta are of an appropriate size, and that there is a balanced level of agreement over all momenta.
In the rest of this section, we provide evidence to explain why the $G^0W^0$ (FSM) prediction is so accurate.

\subsection{Spin moment fixing}
Crucially, we note that the moment for the $G^0W^0$ calculation alone (without FSM) is too high (0.74~$\mu_B$, see Table~\ref{t:mom}) and, correspondingly, the total area of the MCP is badly wrong even though the overall shape is already essentially correct.
We must conclude that $G^0W^0$ alone is not adequate and that to comprehensively describe the magnetization in Ni, something else is essential.
However, since the $G^0W^0$ (FSM) prediction is remarkably good, we find that an effective magnetic field can provide the necessary area reduction \textit{without} reducing the fidelity of the overall shape.
We conclude that the MCPs from $G^0W^0$ with DMFT should also be appropriate because, interestingly, we only find small differences between the DFT (FSM) and the DFT+DMFT MCPs (shown in Fig.~\ref{fig:theory-vs-experiment}).
This agrees well with the conclusions by Sponza {\it et~al.}~\cite{sponza_2017} who showed that 
QSGW calculations gave an enhanced magnetic moment and exchange splitting, which could be brought into agreement with the experimental values by using either DMFT or an external magnetic field.
The magnitude of the DFT+DMFT moment (and exchange splitting) have been found to depend on the DMFT interaction Hamiltonian's parameters $U$ and $J$~\cite{james2021magnetic}, where these parameters influence the DFT+DMFT predicted local electron-correlation-induced temporal spin fluctuations.

We note that both DFT (FSM) and DFT+DMFT do, on their own, improve the agreement compared to DFT by reducing the total area of the MCPs to match the experiment.
We find the problem is that the sizes of specific features (peaks and troughs) are not correct and an imbalance appears so that at low momenta the spin momentum density is overpredicted and at high momenta it is underpredicted.
It was previously confirmed that there is no alternative DMFT $U$ (and $J$) value which can be used to obtain overall agreement with the experimental data over the entire momentum domain~\cite{james2021magnetic}.

Considering the similarity between DMFT and a static effective magnetic field in this case, as well as the fact that one of these is need to correct the exchange splitting (which for $GW$ approximations is the same or even worse than for DFT \cite{aryasetiawan_self-energy_1992,sponza_2017}), we believe that this strongly indicates that local (temporal) spin fluctuations~\cite{halilov2006physics,Shimizu_1981,moriya2000,moriya2006} have a strong impact in Ni. The static effective magnetic field does a good job of mimicking the significant effects the local spin fluctuations have on the magnetization (and exchange splitting) within Ni. 
We emphasise that all of this supports the conclusion that it is difficult to theoretically predict the MCP accurately unless all aspects of the magnetism are well described.
Having established the importance of the FSM component (or the effect of including DMFT) of the $G^0W^0$~(FSM) calculations, we now return to considering why the $G^0W^0$ component is so important.

\subsection{Localized and delocalized magnetism}

Now we will break down the contributions to the MCPs. 
It was shown by looking at the contributions to the MCPs from each band within DFT that there is a negative contribution to the total MCPs from the itinerant $s$- and $p$-like bands within the low momentum region (up to about 1.5 a.u.) with respect to the positive contribution of the $d$-bands~\cite{ku.as.90,ti.br.90}. This negative contribution is referred to as the negative spin polarization. This negative spin polarization is the result of the more itinerant interstitial electrons screening (opposing) the positive local moment which is mainly associated with the tightly bound $d$ electrons. Neutron scattering measurements~\cite{Mook_1966} also confirm this behaviour.

\begin{figure*}[t!]
  \centerline{\includegraphics[width=0.95\linewidth]{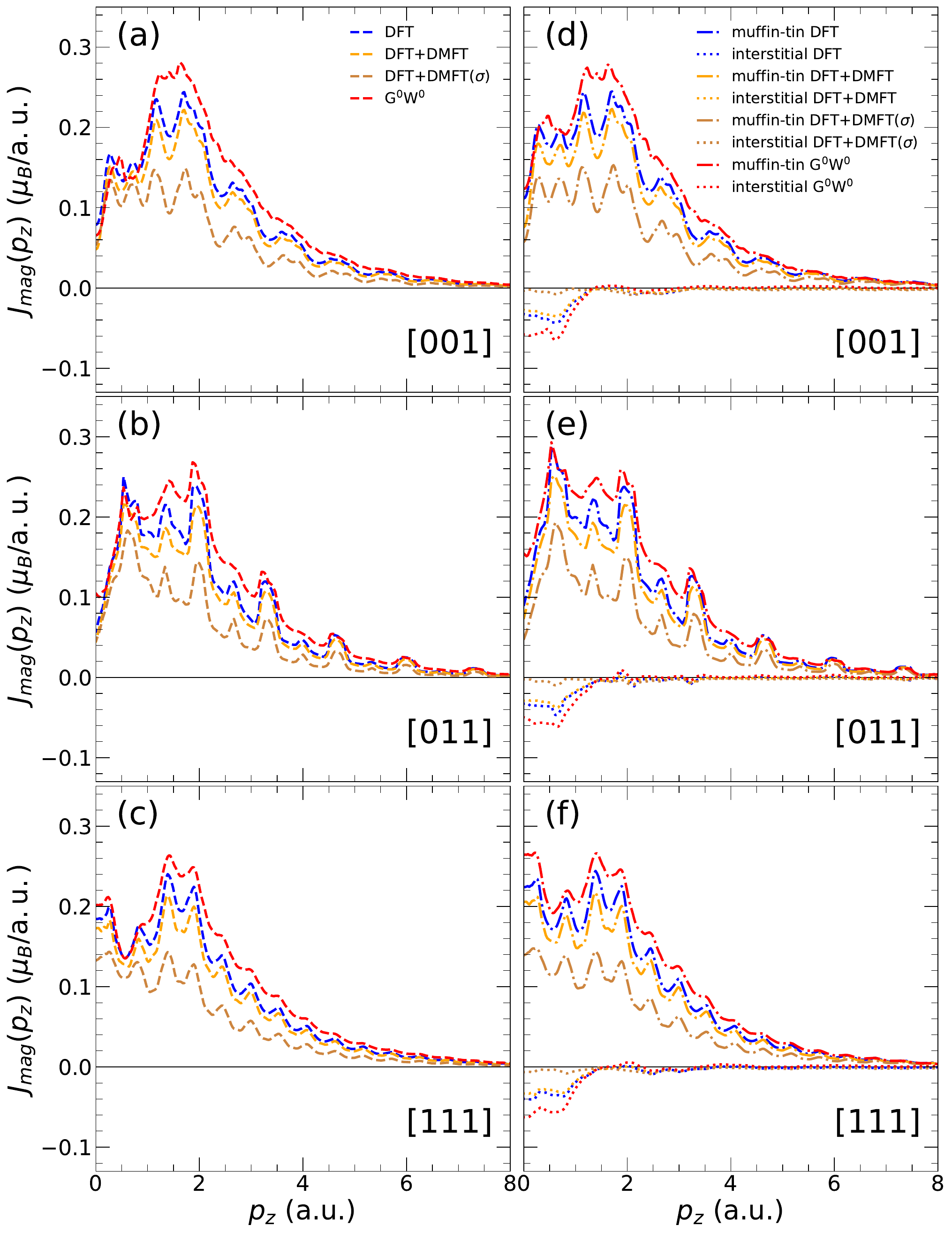}}
  \caption{Comparison of the unconvoluted high symmetry direction ([001], [011], and [111]) MCPs broken down into muffin-tin (near-nuclei) and interstitial region contributions.
  We present DFT (LSDA), DFT+DMFT,  DFT+DMFT($\sigma$) and $G^0W^0$ calculations. 
  In panels (a)-(c) we show the total MCP for each method and high-symmetry direction. 
  In panels (d)-(f) we show the muffin-tin and interstitial basis function resolved MCPs for each high-symmetry direction and method.}
 \label{fig:MT-and-interstitial}
\end{figure*}

We present the muffin-tin (localized) and interstitial (delocalized) contributions to the MCPs in Fig.~\ref{fig:MT-and-interstitial}.
We see that both local (Heisenberg model) magnetization and itinerant magnetization are present here~\cite{Shimizu_1981,moriya2000,moriya2006,kittel2005introduction}. The ferromagnetic order (positive spin momentum density) clearly stems from relatively localized spins (at the atomic-like-character regions near to the nuclei), and in stark contrast the spins which are relatively delocalized and mobile (itinerant) prefer to oppose (screen) the net ferromagnetic alignment, reducing the total spin moment. 
The prominence of the umklapp peaks have been suppressed in the $G^0W^0$ [and $G^0W^0$ (FSM)] muffin-tin MCP contribution. We note that this suppression originates from the smearing of the occupation distribution in the $G^0W^0$ calculation. The umklapp peaks are still visible (especially in the [011] MCP) in both the $G^0W^0$ and experiment~\cite{di.du.98} showing that the smearing of the quasiparticle occupations from correlations do not fully ``wash out'' the umklapp peaks. Furthermore, we find that the $G^0W^0$ quasiparticle many-body occupation distribution helps with improving the total MCP shape, but this is not the primary reason for the excellent agreement between the $G^0W^0$ and experimental MCPs.

The $G^0W^0$ calculation enhances the local magnetization, and thus the total spin moment with respect to the DFT total spin moment. (The spin moment contributions of each theoretical method are given in Table~\ref{t:mom}.)
The (magnitude of the) $G^0W^0$ delocalized magnetization is also enhanced, but it only significantly contributes at very low momenta. This explains the improvement in the shape within the low momentum region of the MCPs. The influence of an effective magnetic field or DMFT within DFT+DMFT, on the other hand, significantly suppresses the muffin-tin spin moment as it reduces the exchange-splitting~\cite{james2021magnetic} by mainly adjusting the $d$-states. Looking at the FSM (and the DFT+DMFT) calculations, the spin moment contributions are somewhat similar across these calculations apart from the $G^0W^0$ (FSM) low momentum interstitial spin moment which is greater than the total interstitial spin moment contribution.  This change in the negative spin polarization from the more delocalized electrons within $G^0W^0$ indicates a change in the screening of the localized magnetization.
We conclude that these changes of both local and non-local types of magnetization is utterly key to balancing the MCPs over high and low momenta and predicting the correct shapes.

The importance of non-local correlations explains why DFT+DMFT cannot predict the MCPs so well \cite{james2021magnetic}.
This is a point that we can now reinforce by taking brief detour to consider the DFT+DMFT($\sigma$) method.
We see that that Ni MCPs from DFT+DMFT($\sigma$) calculations in Fig.~\ref{fig:MT-and-interstitial} are vastly different from the DFT+DMFT MCPs and are in very poor agreement (far worse than the DFT results) when compared with the experimental data (including for different $U$ and $J$ values~\cite{james_phd_2021}). 
In Table~\ref{t:mom}, we see that the spin moment of DFT+DMFT($\sigma$) is in very poor agreement with the experimental one, and the DFT+DMFT($\sigma$) MCPs have no significant negative spin polarization in the interstitial MCPs and are almost entirely described by the muffin-tin MCP contribution.
We note that this emphasises the importance of the non-local correlations and states other than $d$-character states, including those with negative spin polarization, in determining the total magnetization. We emphasise that the importance of DMFT in Ni here is that it includes the local (temporal) spin fluctuations needed to obtain the correct exchange splitting and (spin) moment.

\subsection{Diagonal $G^0W^0$ approximation and other $GW$ approximations}
\label{sec:diagonal_GW}

\begin{figure*}[!t]
  \centerline{\includegraphics[width=0.95\linewidth]{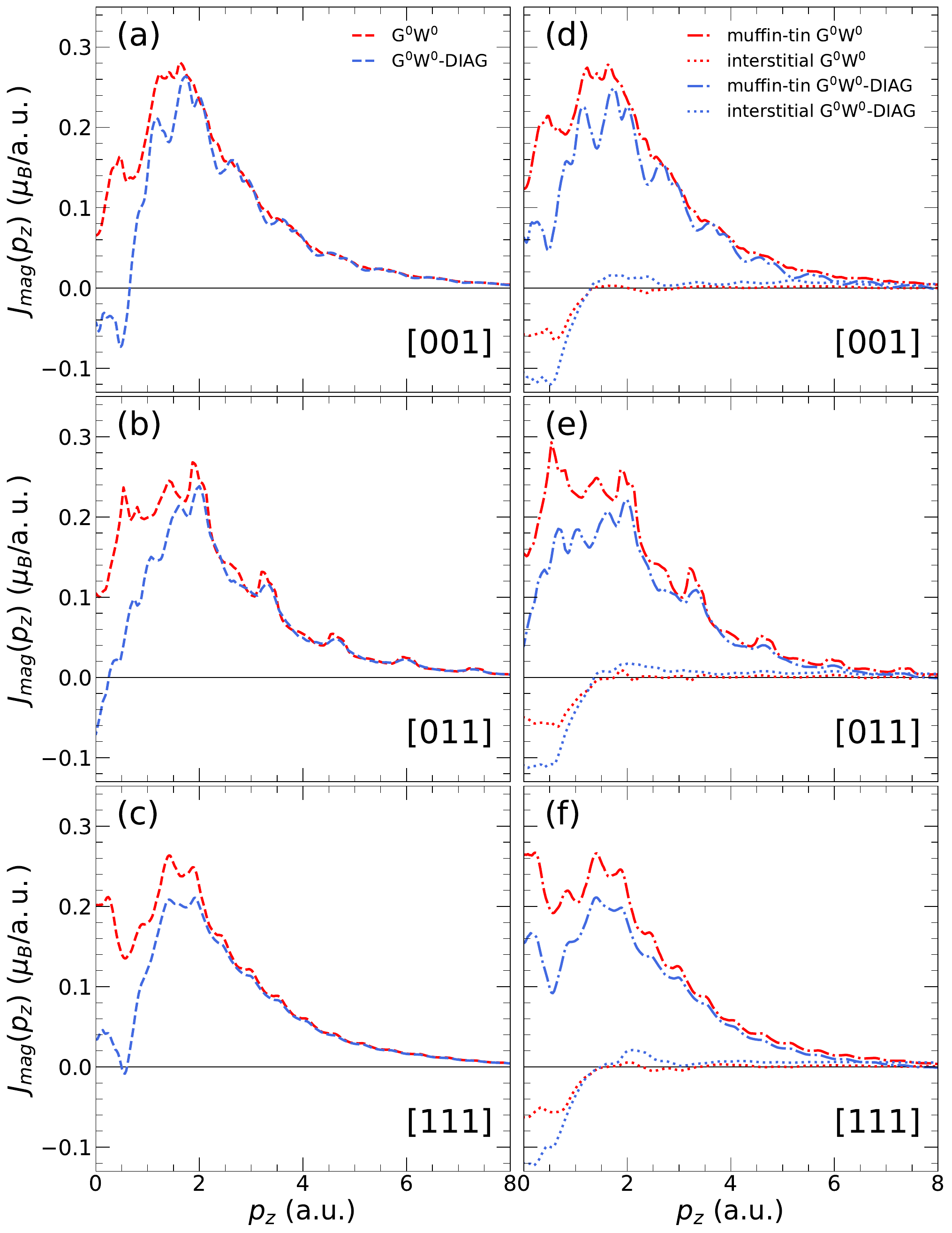}}
  \caption{Comparison between MCPs which we obtain for our $G^0W^0$ approximation and diagonal $G^0W^0$ ($G^0W^0$-DIAG) calculations.
  We present these results without convolution with an experimental resolution function.
  We show the total MCP for each method and high-symmetry direction in panels (a)-(c). 
  We show the muffin-tin and interstitial basis function resolved MCPs for each high-symmetry direction and method in panels (d)-(f).
}
 \label{fig:GW-comp}
\end{figure*}

Here we have used a rather complete version of $G^0W^0$. We solve the Dyson equation, we ensure numerical convergence by using an exceptionally high Matsubara frequency cutoff, and we use the off-diagonal self-energy elements.
However, it is common to assume that the diagonal approximation to the $GW$ self-energy should be adequate (because the computational expense of $GW$ is often otherwise too high).
With Fig.~\ref{fig:GW-comp} we demonstrate that this widely made assumption is seriously flawed in this case, by showing that the spin density in momentum space, and the muffin-tin and interstitial contributions, are described extremely poorly in the diagonal $G^0W^0$ approximation.
The total, muffin-tin and interstitial spin moments also badly fail to match the $G^0W^0$ values (Tab.~\ref{t:mom}).
We believe that this explains why previous $G^0W^0$ calculations were not more accurate than DFT \cite{kubo_electron_2004}.
The diagonal approximation should generally be valid if the full effect of the $G^0W^0$ calculation is to simply rigidly energetically shift the bands from their starting points, without significantly changing the band shapes or the ways that the bands cross over each other.
Clearly, we do not find this condition to be valid for Ni.

Finally, we now consider how the one-shot $G^0W^0$ approximation that we have applied here is appropriate in comparison to other theoretical approximations.
For a normal $GW$ approximation calculation, without a vertex correction, the one-shot method is better justified than a fully self-consistent $GW$ method \cite{hedin_correlation_1999} 
($G^0W^0$ typically predicts experimental results more accurately).
In our view it would be even better to use a quasiparticle selfconsistent method (called QSGW) to find a starting point for the $G^0W^0$ calculation \cite{kotani_quasiparticle_2007,van_schilfgaarde_adequacy_2006,ismail-beigi_justifying_2017}.
Vertex corrections to the $GW$ approximation \cite{kutepov_electronic_2016,cunningham_quasiparticle_2023} may be interesting to test in the future too (these account for a greater range of correlations and can make fully self-consistent calculations more valid).
Our results point to the strength of a unified $GW$+DMFT theoretical approach: Ni MCPs from self-consistent ``GW plus DMFT'' methods \cite{biermann_first-principles_2003,choi_first-principles_2016} could be calculated in the future to confirm the (likely high) level of agreement with the MCP experimental data.

\section{Conclusion}

We conclude that the LSDA spin density in momentum space is flawed in comparison with the experimental data which implies that it is fortuitous that the Ni LSDA total spin moment is somewhat accurate, but not for the right reasons.
Interestingly, the authors of Ref.~\cite{sponza_2017} also reach this conclusion, but via a different and complementary route.
Since various magnetic properties such as the exchange-splitting and total spin moment are all reflected in MCPs, we conclude that these (or more generally the spin density in momentum space) act as a sensitive and complete test of the quality of a theoretical prediction of the magnetization.
Indeed, to predict MCPs that agree with the experimental ones, we provide evidence that several things beyond DFT are needed: a full treatment of non-local correlation to adjust both the local and the itinerant magnetization (balancing the MCP contributions) and treatment of fluctuations of local spin for reducing the  exchange splitting and spin moment (and hence the MCP area).
Based on this work, we strongly encourage magnetic Compton scattering experiments with even higher resolution to be performed for Ni.
Undoubtedly these findings are relevant to other magnetic materials too.

\begin{acknowledgments}
A.D.N.J.~acknowledges funding and support from the Engineering and Physical Sciences Research Council (EPSRC). 
Calculations were performed using the computational facilities of the Advanced Computing Research Centre, University of Bristol (\href{http://bris.ac.uk/acrc/}{http://bris.ac.uk/acrc/}). We appreciate the insightful input and discussions from Dr. J. K. Dewhurst.
\end{acknowledgments}

\bibliography{ref}

\begin{table*}[ht!]
\centering
\begin{tabular}{c|c|c|c|c|c}
& muffin-tin spin & muffin-tin spin & interstitial spin & interstitial spin & total spin \\
& moment $(\mu_{\rm B})$ & moment $\leq$ 1.5 a.u. $(\mu_{\rm B})$ & moment $(\mu_{\rm B})$ & moment $\leq$ 1.5 a.u. $(\mu_{\rm B})$ & moment $(\mu_{\rm B})$ \\
\hline
\hline
DFT & 0.682 & 0.294 & -0.052 & -0.039 & 0.630  \\
DFT (FSM) & 0.607 & 0.260 & -0.041 & -0.035 & 0.566  \\
DFT+DMFT & 0.605  & 0.261 & -0.046 & -0.032  & 0.559   \\
DFT+DMFT($\sigma$) & 0.409 & 0.184 & -0.031 & -0.006  & 0.378   \\
G$^0$W$^0$ & 0.794 & 0.333 & -0.051 & -0.058 & 0.743 \\
G$^0$W$^0$ (FSM) & 0.599 & 0.251 & -0.040 & -0.047 & 0.559 \\
G$^0$W$^0$-DIAG & 0.589 & 0.213 & -0.036 & -0.101 & 0.552  \\
\hline
\end{tabular}
\caption{The total, muffin-tin and interstitial theoretical spin moments from the DFT (LSDA), DFT fixed spin moment (FSM), DFT+DMFT, DFT+DMFT($\sigma$), $G^0W^0$, $G^0W^0$ (FSM) and diagonal self-energy $G^0W^0$ ($G^0W^0$-DIAG) calculations. 
We obtain these spin moments by integrating the MCPs over momentum space.
We also present the spin moments which were obtained by integrating only up to a momentum value of 1.5 a.u. because this low momentum domain is where the relatively delocalized spins contribute.
The experimental spin moment for Ni is 0.56 $(\mu_{\rm B})$~\cite{di.du.98}.
}\label{t:mom} 
\end{table*}

\end{document}